\documentclass[useAMS,usenatbib,usegraphicx]{mn2e}
\usepackage{txfonts}
\DeclareMathAlphabet{\mathpzc}{OT1}{pzc}{m}{it}

\font\sc=cmcsc10
\def\kms{\;{\rm km\,s}^{-1} }
\def\Msun{\,M_\odot}
\def\Lsun{{\rm L}_{\rm \odot}}
\def\vdisp{\langle v^2 \rangle}
\def\sqvdisp{{\langle v^2 \rangle}^{1/2}}
\input epsf

\title[The core of the Sextans dSph]{A photometrically and
kinematically distinct core in the Sextans dwarf spheroidal galaxy}
\author[J.~Kleyna et al.]
{Jan T. Kleyna$^1$, Mark I. Wilkinson$^2$, 
         N. Wyn Evans$^2$, Gerard Gilmore$^2$\\
$^1$Institute for Astronomy, University of Hawaii, 2680
            Woodlawn Drive, Honolulu, HI 96822, USA \\
$^2$Institute of Astronomy, Madingley Road,Cambridge, CB3 OHA, UK.}

\begin{document} 

\maketitle
\begin{abstract} 
We present the line of sight radial velocity dispersion profile of the
Sextans dwarf spheroidal galaxy (dSph), based on a sample of 88 stars
extending to about $1^\circ$ ($\sim 1.5$kpc).  Like the Draco and Ursa
Minor dSphs, Sextans shows some evidence of a fall-off in the velocity
dispersion at large projected radii, with significance $p=0.96$.
Surprisingly, the dispersion at the very centre of Sextans is close to
zero (with significance $p=0.96$). We present evidence which suggests
that this latter change in the stellar kinematics coincides with
changes in the stellar populations within the dSph. We discuss
possible scenarios which could lead to a kinematically and
photometrically distinct population at the centre of Sextans.
\end{abstract}


\section{Introduction}
It has long been known that the velocity dispersions of Local Group
dwarf spheroidals (dSphs) imply mass-to-light ratios $M/L$ of up to
$\sim 100 \Msun/\Lsun$.  Until recently, all estimates of dSph $M/L$
have been based upon a measurement of the central velocity dispersion,
and the assumption of mass follows light.  The advent of datasets of
hundreds of discrete radial velocities of bright stars in the nearby
dSphs has changed all this.  To date, the line-of-sight velocity
dispersion profiles in the Fornax, Draco and Ursa Minor dSphs have
been mapped out right to the optical
edge~\citep{mateo97,kleyna02,wilkinson04}.

In this paper we turn our attention to the Sextans dSph and present
the first results from our new radial velocity survey.  The new data
show an almost flat velocity dispersion to large radii, with some
evidence of a fall-off in the outer parts similar to that seen in
Draco and Ursa Minor~\citep{wilkinson04}. Surprisingly, Sextans
appears to be kinematically cold at the very centre.  This is the
second localised cold population observed in a dSph~\citep{kleyna02},
and it coincides with observable changes in the composition of the
stellar population. We suggest that the nucleus of the Sextans dSph
contains a dynamically separate component, possibly the remains of a
star cluster dragged to the centre by dynamical friction.

\section{OBSERVATIONS}

\subsection{Velocities}
\label{subsec:obs}

We observed Sextans with the multifibre instrument AF2/WYFFOS on the
{\it William Herschel Telescope} on La Palma on 11-16 March
2004. Approximately two of these six nights were lost to weather or
technical problems.  We reduced the spectra using a modified version
of the IRAF WYFRED task, and we obtained the final velocities using
the FXCOR cross-correlation function (CCF) velocity package. From the
point of view of the data reduction, Sextans presents particularly
thorny difficulties. Due to the bulk line of sight velocity of
Sextans, all three Ca triplet lines absorption lines used to measure
the velocity are embedded in bright sky emission lines. Hence it is of
paramount importance to subtract the sky spectrum accurately, and to
distinguish between those correlation peaks caused by stellar
absorption lines and those arising from a random or systematic sky
residual.

Each WYFFOS pointing consisted of approximately 75 fibres allocated to
objects, and an average of 27 sky fibres interleaved among the object
fibres.  After rebinning each spectrum to a common linear dispersion
and flattening with a tungsten lamp spectrum, we constructed a sky
spectrum for each object fibre by median-combining the five sky
spectra closest to the object spectrum on the CCD, thereby minimising
the effects of dispersion-direction PSF variations.  Next, we
subtracted the sky spectra from the object spectra by modelling each
spectrum as the sum of a third-order polynomial plus a multiple of the
sky, and minimising the integrated absolute deviation of the model
from the actual spectrum, excluding those regions known to contain
stellar absorption lines.  Effectively, this procedure subtracts out
the correct amount of sky by minimising the sky line residuals.  As
the final step before computing velocities, we shifted each spectrum
to a common heliocentric velocity frame.

For our previous work on Draco and UMi~\citep{kleyna01, kleyna02,
kleyna03, wilkinson04}, the sky lines were safely separated from the
stellar lines, and it was possible to combine all of the good spectra
for an object into a single spectrum, producing an unambiguous CCF
against a synthetic template.  With Sextans, it was difficult to
distinguish good spectra from bad, so we elected to obtain as many
individual measurements as possible by measuring a velocity with each
of the three Ca triplet lines in each one hour exposure, using FXCOR
in an automated mode, and allowing it to select the highest CCF peak
between -200 and +400 $\kms$.  For each velocity, FXCOR also gives an
uncertainty $\propto (1+R)^{-1}$, where $R$ is the Tonry-Davis $R$
value \citep{tonrydavis79}.

Next, we constructed the following sigma-clipping procedure: for a set
of measured velocities and their nominal uncertainties $\sigma_i$, we
compute the median of the set, discard the farthest outlier (in terms
of $\sigma_i$), and iterate until all of the remaining points are less
than 2.5 $\sigma_i$ from the median.  Then we computed the velocity of
each object in each of lines 1, 2, and 3 by applying the sigma
clipping procedure to all measurements in the line in question.  We
observed that the velocities in line 2 and 3 were consistent, but
generally disagreed with line 1 -- this is not surprising, as line 1
is the weakest line, and is atop the strongest sky line.  Thus we
elected to compute our final best velocity for each object by applying
the sigma clipping procedure to the combined set of velocities from
lines 2 and 3.

To accept a velocity for our subsequent analysis, we required that
there exist at least four velocities, representing at least $2/3$ of
the inputs, which survive the sigma clipping procedure.  In all, 88
Sextans member stars from the combined $2+3$ lines within 30 $\kms$ of
the dSph's median velocity met these criteria.

Our sigma clipping and subsequent error estimates depended on the
validity of the $R$-value velocity uncertainties produced by FXCOR.
Each final velocity, as a Gaussian-weighted average of individual
velocities, had an associated $\chi^2$.  The mean $\chi^2$ per degree
of freedom for all member and non-member stars with good velocities
was 0.93, suggesting that the FXCOR velocities were indeed valid.
Repeating the clipping procedure after rescaling the FXCOR errors by
0.75 and 1.33 produced a $\chi^2$ per degree of freedom that deviated
from the expected value of 1 in the expected directions.  The
smallest $P(\chi^2)$ observed for a Sextans member was 0.004, which
would be expected to occur by chance in a sample of this size a third
of the time.

\begin{figure}
\includegraphics[height=7cm, angle=270]{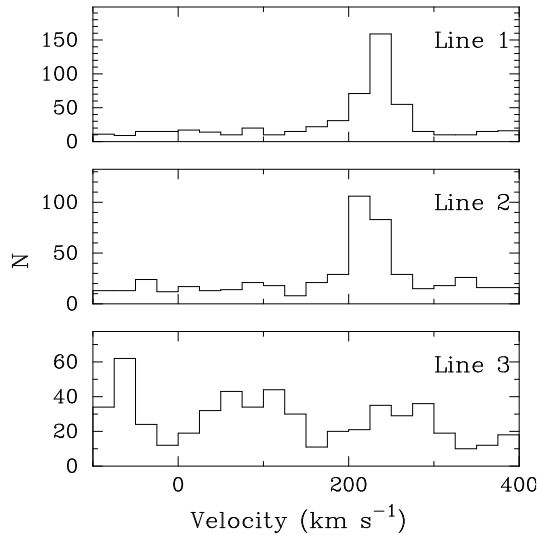}
\caption{Distribution of pseudo-velocities of sky fibres when
processed as though they were data fibres, for each of the three lines
of the Ca triplet.  Lines 1 and 2 produce a spurious peak near the
mean velocity of Sextans, resulting from random and possibly
systematic sky residuals.  Line 3, on the other hand, produces a
relatively unbiased distribution of spurious velocities.  }
\label{fig:skyvel}
\end{figure}

Despite our care to sift out spurious velocities from our data, the
sky line contamination remains a concern.  Fig.~\ref{fig:skyvel} shows
the result of reducing our sky fibres as though they were object
fibres.  Because lines 1 and 2 of the Ca triplet are directly under
two strong sky lines, the residue of the sky tends to produce spurious
velocities at 233 and 223 $\kms$, respectively.  The peaks actually
becomes {\sl stronger} if we limit ourselves to larger Tonry-Davis $R$
values, probably because we are picking out spectra with systematic
sky subtraction problems. Only line 3, embedded in a dense cluster of
lower amplitude sky lines, shows an unbiased distribution of spurious
velocities.  Because of this potential source of contamination, we
will take care to verify any observed kinematical qualities of Sextans
using line 3 alone, as well as with our best $2+3$ velocities.

\begin{figure}
\begin{center}
\includegraphics[height=7cm, angle=270]{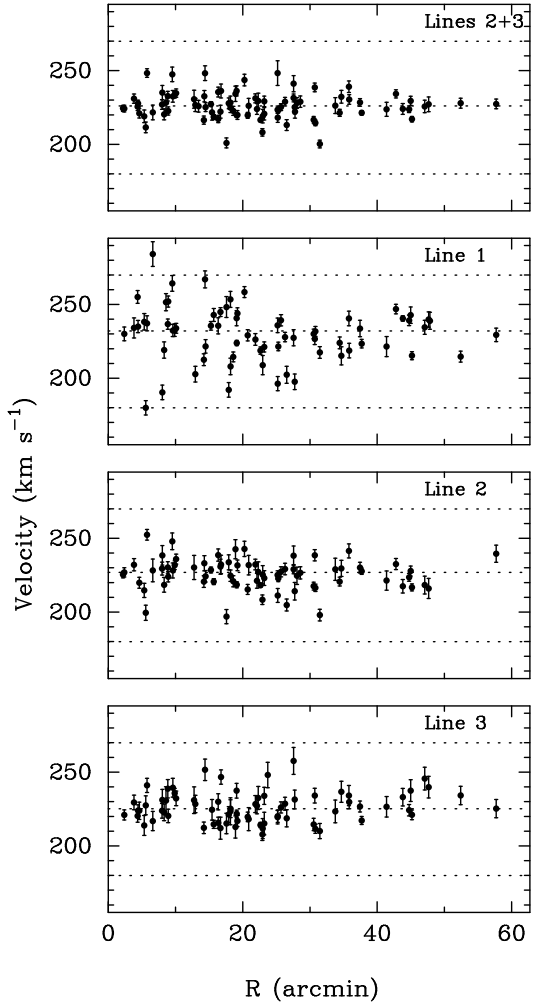}
\end{center}
\caption{Individual velocities as a function of projected radius $R$
for the combined data (using lines 2 and 3 of the Ca triplet), and for
lines 1, 2, and 3 individually.  Outer dotted lines show our
membership limits, and the inner dotted line is the median of the
entire data set.}
\label{fig:velvsr}
\end{figure}

Fig.~\ref{fig:velvsr} shows our individual velocities as a function of
projected radial distance.  The data using only line 1 are obviously
of much poorer quality than for lines 2 and 3; in particular, the
median velocity for line 1 is $232.5\,\kms$, whereas the medians for
lines 2 and 3 are 226.6 and 225.3 $\kms$, respectively.  Immediately
apparent to the naked eye is the apparent tendency of the outermost
stars of the top $2+3$ panel to cluster very tightly around Sextans'
mean velocity; this kinematic coldness at large $R$ was also found to
be present in UMi and possibly Draco \citep{wilkinson04}.
Unfortunately, the data for lines 2 and 3 alone are not of
sufficiently good quality to support this conclusion on their own.

\subsection{Comparison with previous work}
\label{subsec:prev}

In the past, repeated velocity measurements of individual Sextans
stars have been inconsistent from observer to observer, probably
because of the severe sky contamination.  Suntzeff et al. (1993),
(henceforth S93), measured 43 velocities in Sextans using the Argus
multifibre spectrograph on the CTIO 4m; five of these were in the
previous set of Da Costa et al.~ (1991).  On the basis of the
disagreement of the two data sets, S93 argue that Da Costa et
al.~(1991) underestimated their errors.  Similarly, Hargreaves et
al.~(1994), hence H94, observed 26 Sextans spectra using the
single-slit ISIS instrument on the WHT; 11 of these were in the sample
of Suntzeff et al.~(1993).  H94 find that the two data sets disagree,
and conclude that S93 understated their velocity uncertainties. 

Continuing the Sextans tradition of finding fault with our predecessors, we
find that our data agree neither with S93 nor with H94, with whom we
share 16 and 7 stars, respectively.  For the purpose of comparing our
velocities to previous results, we consider only the 38 S93 velocities
without a ``very poor correlation'', and the 21 H94 velocities that
have a Tonry-Davis $R$ above their cutoff of 7.5.  We find a
discrepancy, expressed in $\chi^2$ per degree of freedom, of
$\chi^2/N_{\rm dof}=2.44$ with S93, comparable to the $\chi^2/N_{\rm
  dof}=2.82$ observed between S93 and H94.  With H94, we find a
disagreement of $\chi^2/N_{\rm dof}=7.46$; this value is larger than
the disagreement with S93 probably because the errors of H94 are much
smaller ($2\,\kms$) than those of S93 (median $5.5\, \kms$).

\begin{figure}
\includegraphics[height=7cm, angle=270]{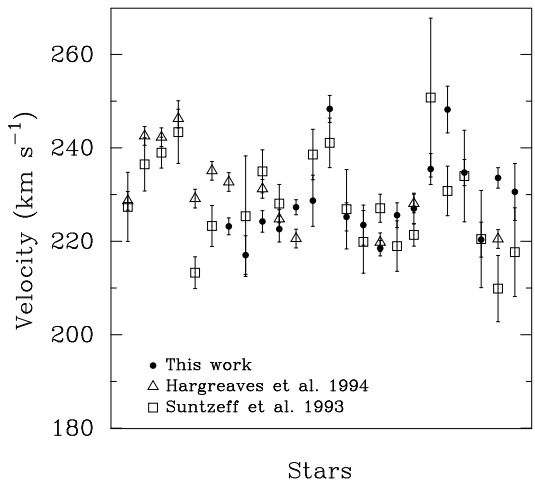}
\caption{Subset of Sextans velocities (with 1-$\sigma$ error bars)
common to at least two of the following: this work, Suntzeff et
al.~(1993), or Hargreaves et al.~(1994).  Each star occupies a unique
position along the horizontal axis, with the two or three different
measurements along the vertical axis.}
\label{fig:hargsuntz}
\end{figure}

Figure \ref{fig:hargsuntz} shows all stars that overlap in at least
two of the data sets.  None of the three data sets is in obvious
systematic disagreement with the other two.  Compared with S93, we
have the advantage of a more stable instrument and a more robust
wavelength solution - S93 needed to adjust their final velocities by a
large systematic factor dependent on the fibre number, whereas our
dispersion solution was stable to $\sim 0.8\,\kms$, verified by
cross-correlating each non sky-subtracted spectrum against a combined
sky template.

Compared with H94, we have the benefit of many more measurements per
star, which we increased by a further factor of three by using each
line separately; this permits testing each star's internal
consistency.  H94, on the other hand, occasionally use only one
measurement per star, relying on a global calibration of error as a
function of Tonry-Davis $R$.  On the other hand, we accept low values
of Tonry-Davis $R$, opting to reject outliers with the procedure
described in \S\ref{subsec:obs}.  As described above, this procedure
gives a final $P(\chi^2)$ very consistent with correct normally
distributed Tonry-Davis $R$ errors. In further Monte-Carlo tests of
Sextans-like data 50\% contaminated with a flat distribution of
spurious velocities, it has proven to be very robust at removing the
outliers and producing data sets with the correct mean and dispersion.
Nevertheless, concern remains that both we and previous observers might be
permitting spurious or sky-shifted correlation peaks into the final
data set.

%%%%%%%%%%%%%%%%%%%%%%%%%%%%%%%%%%%%%%%%%%%%%%%%%%%%%%%%%%%%%%%%

\subsection{Velocity dispersion}
\label{subsec:disp}
For their entire samples, S93 and H94 reported velocity dispersions of
$6.2^{+0.9}_{-0.9}\,\kms$ and $7.0^{+1.3}_{-1.0}\,\kms$, respectively.
However, using their more appropriate maximum likelihood formulation
for the dispersion rather than the method of Armandroff \& Da Costa
(1986), H94 recompute the dispersion of the S93 data to be between 6
and 8 $\kms$, depending on the cut chosen for the Tonry-Davis $R$
value.

H94 express the probability of the line-of-sight velocity dispersion
$\vdisp$ and the mean velocity $\bar v$ for a sample of $N$ velocities
$v_i$ with errors $\sigma_i$ as
\begin{equation}
\label{eq:disp}
%P(\sigma_v,\bar v)\propto \prod_i^N [2\pi (\sigma_i^2 + \sigma_v^2)]^{-1/2} \times
% \exp\left[-{1\over2} {{\left(v_i-\bar v   \right)^2}\over \sigma_i^2 + \sigma_v^2}\right]
P(\vdisp^{1/2},\bar v)\propto \prod_i^N 
{  \exp\left[-{1\over2} {{\left(v_i-\bar v   \right)^2}\over \sigma_i^2 + \vdisp}\right]
   \over
  \sqrt{2\pi (\sigma_i^2 + \vdisp) } }
\end{equation}
They then find the optimal value of $\vdisp$ and $\bar v$ iteratively,
and adjust $\vdisp$ by a factor of $N(N-1)^{-1}$ because the number of
degrees of freedom is reduced by one through the simultaneous
computation of $\bar v$.  Finally, they expand $P(\vdisp^{1/2},\bar v)$
around the optimal point in $\vdisp,\bar v$ to obtain the errors.

We note, however, that Eq.~\ref{eq:disp} is a Gaussian in $\bar v$, so
it can be integrated analytically over $\bar v$ to obtain a one
dimensional probability $P(\vdisp^{1/2})$:
\begin{equation}
\label{eq:sigprob}
P(\vdisp^{1/2}) \propto 
  \sqrt{2\pi \over a} \times
{   \exp[{-{1\over 2} (c-b^2/a) }] \over
   \prod_i^N  {2\pi (\sigma_i^2 + \vdisp) }^{1/2} }
\end{equation}
where $a=\sum_i (\vdisp+\sigma_i^2)^{-1}$, $b=\sum_i v_i
(\vdisp+\sigma_i^2)^{-1}$, and $c=\sum_i v_i^2
(\vdisp+\sigma_i^2)^{-1}$.  $P(\vdisp)$ can then be integrated
numerically to obtain precise error bounds.  This is the approach
taken in this paper: when collectively computing the dispersion of an
entire data set, we use the $P(\vdisp)$ obtained from
Eq.~\ref{eq:sigprob}, effectively integrating over all values of $\bar
v$.  However, when binning the data radially and comparing dispersions
among bins, we wish to use a single mean velocity across all bins, so
we instead compute $P(\vdisp|\bar v)$ by fixing $\bar v$ in
Eq.~\ref{eq:disp} at the median velocity of our entire data set.  In
these cases, we may introduce the further modification that the
Gaussian in Eq.~\ref{eq:disp} is replaced by a Gaussian convolved with
a binary velocity distribution, as described in \cite{kleyna02}.  To
ensure that our implementations of Eq.~\ref{eq:disp} and
Eq.~\ref{eq:sigprob} are correct, we have conducted extensive
Monte-Carlo tests using artificial data.

We note that the data of S93 are contained entirely within the central
$16.35^\prime$, whereas the velocities of H94 extend outside this
radius.  For consistency, we limit our comparisons among different
observers' dispersions to this common central region.  Using our
global, purely Gaussian $P(\vdisp)$, we compute the following most
likely dispersions inside $16.35^\prime$, with central $68\%$
confidence intervals -- this paper:
$\vdisp^{1/2}=7.80^{+1.72}_{-0.94}\,\kms$ from 30 stars; S93:
$7.74^{+1.92}_{-1.17}\,\kms$ from 38 stars; H94:
$8.62^{+3.03}_{-1.10}\,\kms$ from 14 stars. These dispersions are
somewhat larger than the 6 to 7 $\,\kms$ dispersions given by S93 and
H94.  Nevertheless, they are statistically consistent, and the
disparities probably arise from the different subsamples used, and the
slightly different (and perhaps more rigorous) method we use to
compute the dispersion.  We also exclude the low-dispersion stars of
H94 outside $16.35^\prime$, increasing the dispersion we measure in
the H94 data.  Given the disagreements we noted among the three data
sets, it is reassuring that they give such similar dispersions.  This
agreement may arise from the fact that the dispersion is sensitive to
excess velocity errors only when these errors become large compared to
the dSph's overall dispersion, whereas the $\chi^2$ disagreement among
data sets becomes large when the excess velocity errors are large
compared to the nominal velocity uncertainties.

\begin{figure}
\includegraphics[height=7cm, angle=270]{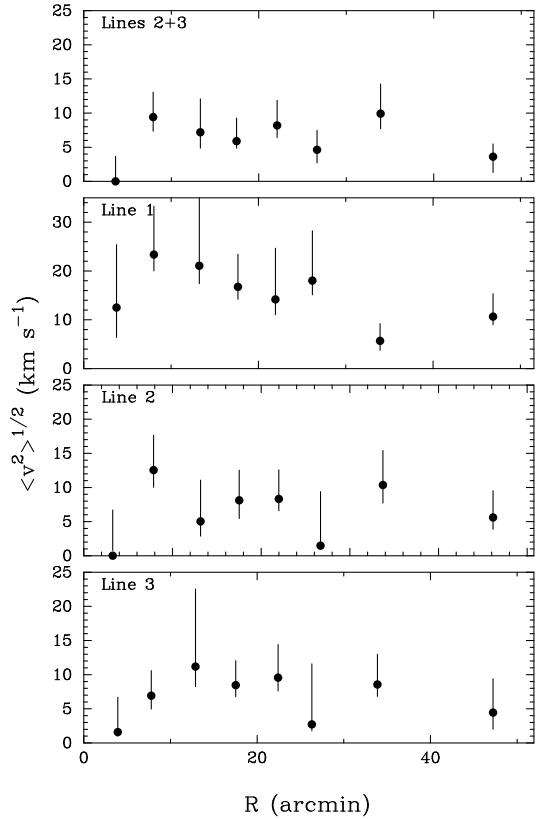}
\caption{Velocity dispersions inside bins with boundaries at projected
radius $R=5^\prime$, $10^\prime$, $15^\prime$, $20^\prime$,
$25^\prime$, $30^\prime$, and $40^\prime$.  The top panel has
velocities computed using both Ca triplet lines 2 and 3, and the other
panels are for lines 1, 2, and 3 individually.  Velocities using line
1 are of poor quality and give much larger dispersions than velocities
obtained with lines 2 and 3.  A binary fraction of 40\% is assumed
\citep{kleyna02}.  Each point represents the most likely dispersion,
and the error bar is the central 68\% confidence interval.  The $R$
value of each point is the mean projected radius of the stars in the
corresponding bin.  The three good data sets are suggestive of a
dispersion that is zero in the centre, rises to about $8\,\kms$, and
then falls off at large radii.}
\label{fig:disp}
\end{figure}
\begin{figure}
\begin{center}
\includegraphics[height=6cm, angle=270]{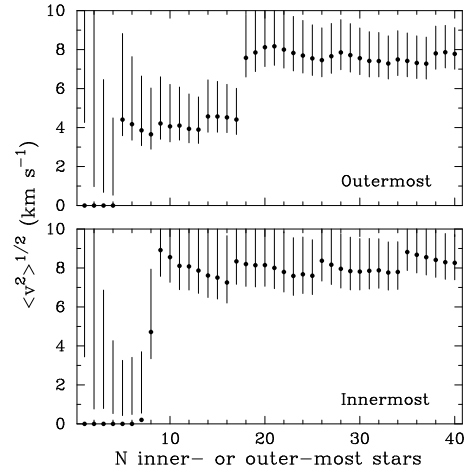}
\end{center}
\caption{Velocity dispersion computed using the $N=1\ldots 40$ innermost or 
outermost stars, for our combined $2+3$ dataset.  The point at $N$
represents the most likely dispersion for the sample of stars from $1$
to $N$ using the purely Gaussian formulation of Eq.~\ref{eq:disp} and
setting $\bar v$ equal to the median of our entire sample. The bars
are the central 68\% confidence intervals at each point.  The bars
need not overlap the most likely point if the probability profile is
asymmetric.}
\label{fig:dispinout}
\end{figure}

\begin{figure}
\begin{center}
\includegraphics[height=6cm, angle=270]{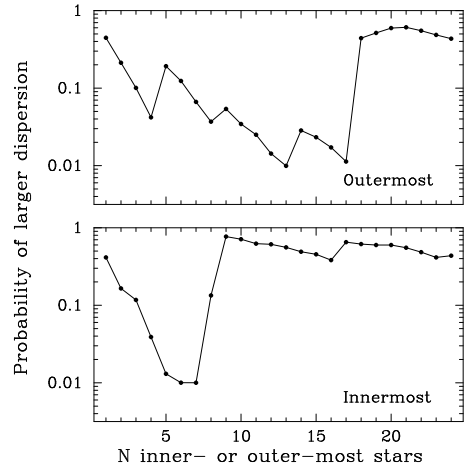}
\end{center}
\caption{Probability that the dispersion in the $N$ innermost or
outermost stars is larger than the dispersion of the entire 88 star
$2+3$ sample, computed according to Eq.~\ref{eq:vdispdifprob}.  There
is $\gtrsim 2\sigma$ evidence of a dispersion drop in both the centre
and outer limits of Sextans.}
\label{fig:dispinoutprob}
\end{figure}

Fig.~\ref{fig:disp} shows the final velocity dispersions as a function
of projected radius for our data set. From Fig.~\ref{fig:disp}, it is
difficult to evaluate whether the statistical significance of the
apparent fall in the dispersion at small and large $R$ is an artefact
of the adopted binning.  Accordingly, in Fig.~\ref{fig:dispinout} we
plot the velocity dispersion advancing from the outermost or innermost
star, each time adding one star to the sample.  We find that the most
likely dispersion for the four outermost and seven innermost stars is
zero.  As we move inward or outward adding stars to the sample, the
upper bound on the dispersion falls until we reach a region of higher
dispersion.  Next, we compute the probability that the inner and outer
samples have a dispersion smaller than the entire ensemble of Sextans
stars as follows: if we have two samples of data $a$ and $b$ with a
common mean $\bar v$, the joint probability distribution for their
dispersions is
\begin{equation}
\label{eq:vdisp2d}
P(\sqvdisp_a,\sqvdisp_b)=\int P(\sqvdisp_a,\bar v)\times P(\sqvdisp_b, \bar v) \, d\,{\bar v}
\end{equation}
where $P(\sqvdisp_a,\bar v)$ is given in Eq.~\ref{eq:disp}.
The probability that $\sqvdisp_a > \sqvdisp_b$  is then
\begin{equation}
\label{eq:vdispdifprob}
P(\sqvdisp_a\!>\!\sqvdisp_b)\!=\!\int
\displaylimits_{\sqvdisp_a\!>\!\sqvdisp_b}
\!P(\sqvdisp_a , \sqvdisp_b) \!d\sqvdisp_a\!d\sqvdisp_b
\end{equation}
This is accomplished identically to the step that converts
Eq.~\ref{eq:disp} to Eq.~\ref{eq:sigprob}.
Fig.~\ref{fig:dispinoutprob} shows the probability, using
Eq.~\ref{eq:vdispdifprob}, that the $N$ innermost or outermost stars
have a velocity dispersion smaller than that of the rest of the
sample.  We find that the 17 outermost stars of Sextans are
kinematically colder than the rest of the dSph at up to the
$p=0.99$ confidence level.  We can draw the cutoff almost anywhere
between 4 and 17 and obtain the same result, so this conclusion is not
finely tuned to the value of our one free parameter, the cutoff
radius.  Similarly, the innermost 7 stars are also cold at the $p=0.99$
confidence level.  In this case, the choice of cutoffs is narrower,
with only 4 to 7 giving this level of confidence.  We also performed a
Monte-Carlo test using simulated data with a flat $8\,\kms$ dispersion
and our nominal observational errors to determine how frequently we
would expect to observe dips of the magnitude seen in
Fig.~\ref{fig:dispinoutprob} {\sl anywhere} in the first or last
$N=24$ stars -- we find that the lowest outer and inner
dispersion dips are significant at the $p=0.96$ level.  Recreating
Fig.~\ref{fig:dispinoutprob} for Ca triplet lines 2 and 3 taken alone
gives similar dips, but at lower significance levels $p=0.8$ to
$p=0.95$.

%% Smallest prob [Innermost] is 0.009223628307755128d0
%% Smallest prob [Outermost] is 0.008977346342644337d0

\begin{figure}
\begin{center}
\includegraphics[width=0.3\textwidth]{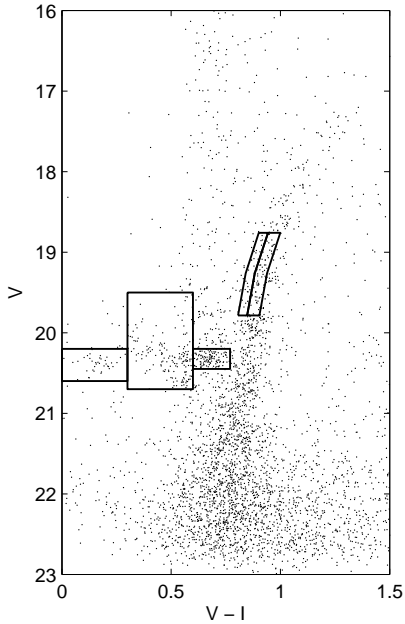}
\end{center}
\caption{$V,V-I$ colour-magnitude diagram of stars within 20 arcmin of
the centre of Sextans (based on archival INT images). Over-plotted are
the 5 bins used to define the stellar sub-populations. See text for a
detailed discussion.}
\label{fig:CMD_bins}
\end{figure}

\begin{figure}
\begin{center}
\includegraphics[width=0.4\textwidth]{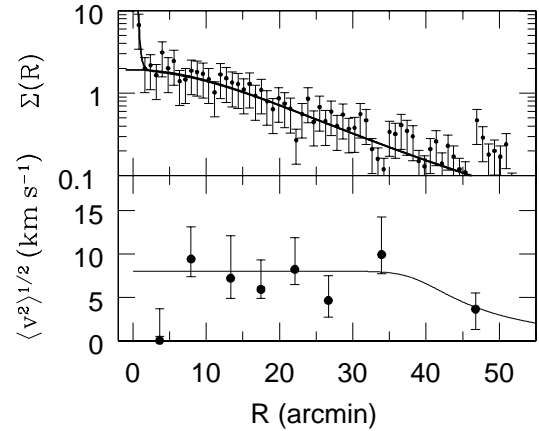}
\end{center}
\caption{{\bf Top:} Surface brightness profile for Sextans together
with the best-fitting single-component Plummer law (lower curve) and a
Plummer law with a central power-law cusp of index $4.4$ (upper
curve). The two curves deviate only inside about 1.6 arcmin where the
cusped model rises to fit the innermost data point. Observed data
taken from \protect\cite{IrHatz95}. {\bf Bottom:} Velocity dispersion
profile with best-fitting model (central data point omitted from the
fit).}
\label{fig:disp_fits}
\end{figure}

\begin{figure}
\begin{center}
\includegraphics[width=0.45\textwidth]{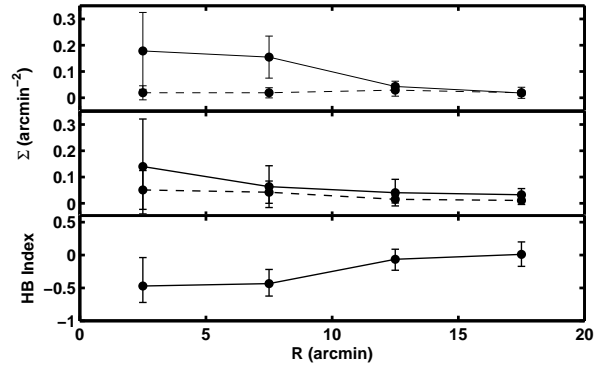}
\end{center}
\caption{Background-subtracted projected density profiles of the red
(solid curve) and blue (broken curve) horizontal branch stars
(top) and of the red (solid curve) and blue (broken curve)
RGB stars (middle) in Sextans based on INT photometric data. The
bottom panel shows the radial variation of the HB morphology
index. All three panels show that the composition of the stellar
population at the centre of Sextans varies with radius on a scale of
about 10 arcmin. See text for a detailed discussion.}
\label{fig:pop_profiles}
\end{figure}

\section{Modelling}

One possible explanation for the observed data would be the presence of a
kinematically distinct stellar component at the centre of Sextans. To test
this idea further, we use the procedure described in~\cite{kleyna03} to
examine the position-velocity data set for signs of substructure. We scan
the face of the dSph with an aperture of radius $5^\prime$. At each
aperture location, we calculate the likelihood that these velocities are
drawn from (i) a single Gaussian with dispersion $\vdisp=7.97\kms$ (ii) a
velocity distribution in which a fraction $f$ of the stars are drawn from
the main Gaussian and the remainder belong to a kinematically distinct
population with different mean velocity and dispersion. Only in an
aperture centred close to the origin does any model other than the main
Gaussian represent a significantly better fit to the velocity data. A
model in which all the stars in this aperture belong to a population with
velocity dispersion $0.5\kms$ (and a mean velocity within $2\kms$ of the
bulk motion of the dSph) is about $195$ times more likely than a model in
which they belong to the main population. Monte Carlo realisations of our
data set demonstrate that the likelihood of observing a spurious cold
clump with the same probability ratio at any location in the cluster is
relatively high. However, only $4.6$ percent of realisations display a
clump with an offset in mean velocity of less than $5\kms$. We therefore
conclude that the central region shows evidence for the presence of a
kinematically distinct population at about the $2\sigma$ confidence
level.

The light distribution of Sextans displays a sharp central
rise~\citep[see][ and the top panel of
Fig.~\ref{fig:disp_fits}]{IrHatz95}. It is tempting to imagine an
association between this feature and the cold population. However, the
difference in the length scales of the two features means that a
direct association is very unlikely. At a radius of 2 arcmin (the
radius of our innermost velocity data point) the ratio of the central
component to the main component is already less than unity, for any
plausible power-law fit to the centre of the light distribution. Thus,
if the central density profile were representative of the spatial
distribution of the cold population, we would not expect that six of
the innermost eight stars in our velocity sample would be cold.

A number of authors have presented evidence for the presence of radial
stellar population gradients in Sextans~\citep{harbeck01,lee03}. In
particular, the red horizontal branch (RHB) stars appear to be more
centrally concentrated than the blue horizontal branch (BHB)
stars. Following these authors, we define five bins in the ($V,V-I$)
colour magnitude diagram (CMD) obtained from deep ($V<22.5$) INT
imaging of Sextans (see Fig.~\ref{fig:CMD_bins}). There are three bins
on the horizontal branch (RHB, BHB, RRLyrae) and two bins on the red
giant branch (RGB) between V=18.8 and V=19.8. The red RGB bin (RRGB)
traces the mean colour of the RGB while the blue bin (BRGB) covers the
blue side of the RGB. The number of stars in each bin is corrected for
foreground contamination by counting the number of stars in a bin of
equal area displaced away from the CMD features associated with
Sextans.

In the top panels of Figure~\ref{fig:pop_profiles} we compare the
background-subtracted surface density profiles of the red and blue
components of the HB and RGB. In agreement with~\cite{harbeck01}
and~\cite{lee03}, we find that the RHB is significantly more centrally
concentrated than the BHB. There is also weak evidence that the RRGB
is more concentrated than the BRGB. The bottom panel of the Figure
shows the radial variation of the HB morphology index $(n_{\rm
BHB}-n_{\rm RHB})/(n_{\rm BHB}+n_{\rm RHB}+n_{\rm RRL})$, where
$n_{\rm BHB}$, $n_{\rm RHB}$ and $n_{\rm RRL}$ are the numbers of BHB,
RHB and RRLyrae stars, respectively~\citep{lee90}. Within 10 arcmin,
the HB is dominated by its red component, while at larger radii, the
numbers of red and blue HB stars are comparable. Given the
uncertainties in the definitions of the CMD bins in our analysis and
the unknown level of contamination of the RHB bins by RRLyraes, the
details of these radial variations are not well determined. However,
it is clear that the composition of the stellar population at the
centre of Sextans varies on a scale of about $5-10$ arcmin.

If both the stellar components of Sextans are tracer populations in a
dark matter dominated potential well, then the more concentrated
population should have a smaller velocity dispersion. The RGBs of the
two populations may be coincident in the CMD and so we would not
necessarily expect to see correlations in our velocity data with
position on the RGB. However, we note that in the innermost radial
bin, the surface density of RHB stars is a factor of 9 higher than
that of the BHB stars. It is therefore plausible that within 5 arcmin
of the centre we would observe mostly stars associated with the cold
component.

One possible origin for the cold, inner population of Sextans is the
preferential retention of star-forming gas at the centre of the dSph
leading to a younger and/or more metal-rich, centrally concentrated
population~\citep{harbeck01}. An alternative explanation is that the
cold central population originated in a star cluster which spiralled
to the centre of Sextans under the influence of dynamical
friction. During the final stages of this process, the cluster
deposited stars in the inner regions of the dSph. It is possible that
the central cusp in the light distribution is the final, probably
unbound, remnant of this cluster. A similar scenario has been proposed
by~\cite{ohlin} for the origin of nucleated dwarf galaxies.

In addition to providing an explanation for the photometric and
kinematic features discussed already, the cluster scenario naturally
accounts for the distribution of blue stragglers in
Sextans. \cite{lee03} noted that the brighter (i.e. more massive) blue
stragglers are more centrally concentrated than the fainter ones. As
those authors comment, mass-dependent spatial distributions of stars
are a generic outcome of mass segregation in stellar systems. However,
the time scale for such segregation to occur in Sextans is very long
due to the low surface brightness of the stellar population. If a
significant fraction of the blue stragglers were formed in a star
cluster which subsequently disrupted near the centre of Sextans, mass
segregation within the cluster would ensure that the most massive blue
stragglers would be the last to be tidally removed from the cluster
and hence would have a more concentrated spatial distribution.

If we assume that all the light in the central 10 arcmin is due to
cluster debris (as suggested by the dominance of the cold population
in the inner regions and the large ratio of RHB to BHB stars inside 10
arcmin) and that the (constant) luminosity density within 10 arcmin is
$0.002$L$_\odot\,$pc$^{-3}$~\citep{mateo98}, we obtain a luminosity of
$1.3\times10^5$L$_\odot$ for the putative cluster. If the mass to
light ratio is about 2 solar units (typical of Galactic globular
clusters), then the mass of the cluster is about
$3\times10^5$L$_\odot$. This is comparable to the masses of Galactic
globular clusters.

We determine the mass profile of Sextans in the context of the cluster
model by assuming a Plummer model for the light distribution,
excluding the central point (see Figure~\ref{fig:disp_fits}).
Similarly, we remove the central point from the velocity dispersion
profile and fit a profile of the form (see Figure~\ref{fig:disp_fits})
\begin{equation}
\label{eq:plum_disp}
\vdisp(R) = \frac{\vdisp_0}{(1 + (R/R_{\rm
c})^\alpha)^{\gamma/\alpha}}
\end{equation}
If we further assume isotropy of the velocity dispersion tensor, then
the mass within 1 kpc (using the Jeans equation) lies in the range
$3\times 10^7 - 1.5\times 10^8$M$_\odot$ -- the exact value is
sensitive to the form assumed for the outer fall-off in the
dispersion. Assuming that the dark matter has an isothermal
distribution out to 1 kpc, then the time taken for a cluster of mass
$3\times 10^5$M$_\odot$ to spiral to the centre due to dynamical
friction would be in the range 0.7-1.5 Gyr~\citep{BT87}. Thus, we
would expect to find any such cluster at the centre of the Sextans
dSph.

We also note that the mass within 10 arcmin ($=250\,$pc) in this model
is approximately $2.5\times10^6$M$_\odot$. Hence, the implied mass to
light ratio of the inner regions is roughly 19 solar units. This is
much higher than the M/L expected for an old stellar population based
on observations of globular
clusters~\citep[e.g.][]{parmgil01,fgw99}. This suggests that the inner
regions of Sextans are dark matter dominated, which is consistent with
our earlier assumption that the stellar populations in the central
regions are not currently self-gravitating.

\section{Conclusions} 

We have presented the radial variation of the velocity dispersion for
the Sextans dSph. The behaviour of the velocity dispersion profile at
large radii is qualitatively similar to that seen in the Draco and
Ursa Minor dSphs~\citep{wilkinson04}. In those galaxies, the fall-off
in the velocity dispersion at large radii coincided with a break in
the light distribution and was interpreted as indicative of mild tidal
perturbation of the stars. In Sextans, the photometry presented
by~\cite{IrHatz95} is of insufficient quality to identify features at
$40-50$ arcmin. The interpretation of the fall-off in the outer
velocity dispersion of Sextans is unclear, but may be due to the tidal
field of the Milky Way.

Intriguingly, the dispersion at the very centre of Sextans is close to
zero and this low value is coincident with significant radial
gradients in the stellar populations. We suggest that this is caused
by the sinking and gradual dissolution of a star cluster at the centre
of Sextans. This is consistent with both our inferred mass for Sextans
and the estimated timescale for dynamical friction to bring any star
cluster to the centre of the Sextans. The model also provides an
explanation for the origin of the central cusp in the light
distribution and the different spatial distributions of the bright and
faint blue stragglers in Sextans. Thus, our hypothesis provides a
reasonable explanation of all the observed data, both kinematic and
photometric, on the Sextans dSph.

\section*{Acknowledgements}
JTK gratefully acknowledges the support of the Beatrice Watson Parrent
fellowship.  MIW acknowledges the support of PPARC.  We are very
grateful to Mike Irwin for providing the astrometric catalogue used to
produce our target list. We thank George Seabroke and the staff at the
Isaac Newton Group at La Palma for their assistance in obtaining the
data for this paper and Dougal Mackey for useful discussions.   We are
grateful to the referee Carlton Pryor for useful suggestions.

{}

\end{document}